\documentclass{article}
\usepackage{arxiv}

\usepackage[utf8]{inputenc} 
\usepackage[T1]{fontenc}    
\usepackage{hyperref}       
\usepackage{url}            
\usepackage{booktabs}       
\usepackage{amsfonts}       
\usepackage{nicefrac}       
\usepackage{microtype}      
\usepackage{lipsum}		
\usepackage{graphicx}
\usepackage{doi}

\usepackage{graphicx}
\usepackage{amssymb, amsmath}
\usepackage{color}
\usepackage{url}
\usepackage{comment}
\usepackage{afterpage}
\usepackage{here}
\newcommand{\reffig}[1]{Figure~\ref{#1}}

\newcommand{\reftab}[1]{Table~\ref{#1}}

\newcommand{\bi}[1]{\ensuremath{\boldsymbol{#1}}}

\title{Airfoil generation and feature extraction using the  conditional VAE-WGAN-gp}

\author{ \href{https://orcid.org/0000-0002-1955-069X}{\includegraphics[scale=0.06]{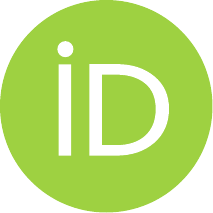}\hspace{1mm}Kazuo Yonekura}\\
	The University of Tokyo,\\
	7-3-1 Hongo, Bunkyo-ku, Tokyo, Japan 113-8656 \\
	\texttt{yonekura@struct.t.u-tokyo.ac.jp} \\
	\And
		Yuki Tomori \\
	The University of Tokyo,\\
	7-3-1 Hongo, Bunkyo-ku, Tokyo, Japan 113-8656 \\
	\And
		Katsuyuki Suzuki \\
	The University of Tokyo,\\
	7-3-1 Hongo, Bunkyo-ku, Tokyo, Japan 113-8656 \\
}

\renewcommand{\shorttitle}{\textit{arXiv} Template}

\hypersetup{
pdftitle={},
pdfsubject={},
pdfauthor={},
pdfkeywords={},
}

\begin{document}
\maketitle

\begin{abstract}
A machine learning method was applied to  solve an inverse airfoil design problem. 
A conditional VAE-WGAN-gp  model, which couples the conditional variational autoencoder (VAE) and Wasserstein generative adversarial network with gradient penalty (WGAN-gp), is proposed for an airfoil generation method, and then it is compared with the WGAN-gp and VAE models. 
The VAEGAN model couples the VAE and GAN models, which enables feature extraction in the  GAN models. 
In airfoil generation tasks, to generate airfoil shapes that satisfy lift coefficient requirements, it is known that VAE outperforms WGAN-gp with respect to the accuracy of the reproduction of the lift coefficient, whereas GAN outperforms VAE with respect to the smoothness and variations of generated shapes. 
In this study, VAE-WGAN-gp demonstrated a good performance in all three aspects. 
Latent distribution was also studied to compare the feature extraction ability of the proposed method. 
\end{abstract}


	\section{Introduction}
\label{sec:intro}
In mechanical design, it is desirable to design shapes that indicate the required performance. 
This task is called an inverse design problem.  
Recently, deep generative models such as generative adversarial networks (GANs) and variational autoencoders (VAEs) have been used for inverse designs \cite{Yonekura21a,Yonekura22SVAE}. 
Both GAN and VAE can generate desired shapes that satisfy these requirements \cite{Yonekura22WGAN}. 
In VAEs, designers can choose the shapes to generate by analyzing latent vectors \cite{Yonekura22SVAE}, which is difficult in GAN. 
In this study,  to enable latent engineering in GAN models, a variational autoencoder generative adversarial network (VAEGAN)  model, a combination of VAE and GAN, is used for inverse designs. 

One approach to an inverse problem and optimization is to use a surrogate model \cite{Braconnier11,Mack2007}. 
The surrogate model is constructed using data, and it outputs the performance parameters from the input shapes in a short time. 
Hence, when a certain shape indicates a certain performance requirement, designers can search from the surrogate models. 
The surrogate model is used for inverse problems and optimization in industrial applications such as turbine cooling holes \cite{Nita14}. 
Deep neural networks are also used as surrogate models to predict performance \cite{Pfrommer18,Liang18}. 
However, the shapes obtained from surrogate models are like those obtained from a training dataset, and completely different shapes cannot be obtained.  

Deep generative models have recently been used to solve inverse problems. 
The GAN \cite{Goodfellow14} is a widely used generative model. 
The airfoil design task is a common target for inverse designs \cite{chen2020airfoil,Sekar19,Yilmaz20}, 
and it is also used in industrial designs \cite{Li21_SP,Wu16,Li2019}. 
In airfoil designs, because the aim is to obtain shapes that satisfy specific aerodynamic requirements, it is important to obtain a smooth shape such that a numerical analysis, such as fluid analysis, can be conducted. 
For this purpose, a conditional GAN is used for the airfoil inverse design \cite{WANG22,Achour20}. 
However, one of the issues with GAN models in airfoil generation is that the output shapes are not smooth, and fluid analysis is not applicable. 
To overcome this issue, a conditional Wasserstein GAN with a gradient penalty (cWGAN-gp) \cite{Arjovsky17b,Gulrajani} is employed, and it outputs a smoother airfoil than the ordinary GAN models as reported in \cite{Yonekura22WGAN}. 
However, GAN models generate data from random latent vectors, and the output shape is chosen randomly. 

VAE \cite{Kingma13} is another generative model. It is used to generate three-dimensional models \cite{Nash17,Guan20} and the airfoil  \cite{Yonekura21a}. 
The VAE is also used for airfoil generation tasks; \cite{Yonekura22SVAE} reported that two different types of shapes are mixed when using VAE models, and such mixing is not possible in the GAN models. 
One of the differences between the GAN and VAE lies in the latent space.
In the GAN model, the generator generates data from a random latent vector or noise vector, whereas in the VAE model, feature vectors are extracted from the training dataset, and data are embedded in the latent space. 
Hence, in the GAN model, the output shape is chosen randomly using the random noise vector. 

VAEGAN \cite{VAEGAN} is a GAN-based neural network that uses a VAE model as its generator. 
The VAE part of the VAEGAN enables feature extraction from data, whereas the entire model is based on the GAN architecture. 
The VAEGAN was applied to the airfoil generation task in \cite{WANG22}, but although the latent space of the VAEGAN is essentially different from that of  the GAN, the latent space has not been discussed in airfoil generation tasks. 
In this study, the VAEGAN was coupled with cWGAN-gp to increase the smoothness of shapes, and it was then applied for the airfoil generation task. The latent space was compared with the cWGAN-gp model, from the comparison, the latent space is well-organized. 

The remainder of this paper is organized as follows:
Section 2 introduces the GAN, VAE, and VAEGAN models. 
The conditional VAEGAN model with the Wasserstein distance and gradient penalty for airfoil generation is proposed in section 3. 
Numerical experiments are presented in section 4, and conclusions are presented in section 5.

\section{Conditional GAN, VAE, and VAEGAN models}
\subsection{Conditional GAN and WGAN-gp}
\begin{figure}[tbph]
	\begin{center}
		\begin{minipage}[h]{\textwidth}
			\begin{center}
				\includegraphics[width=\textwidth]{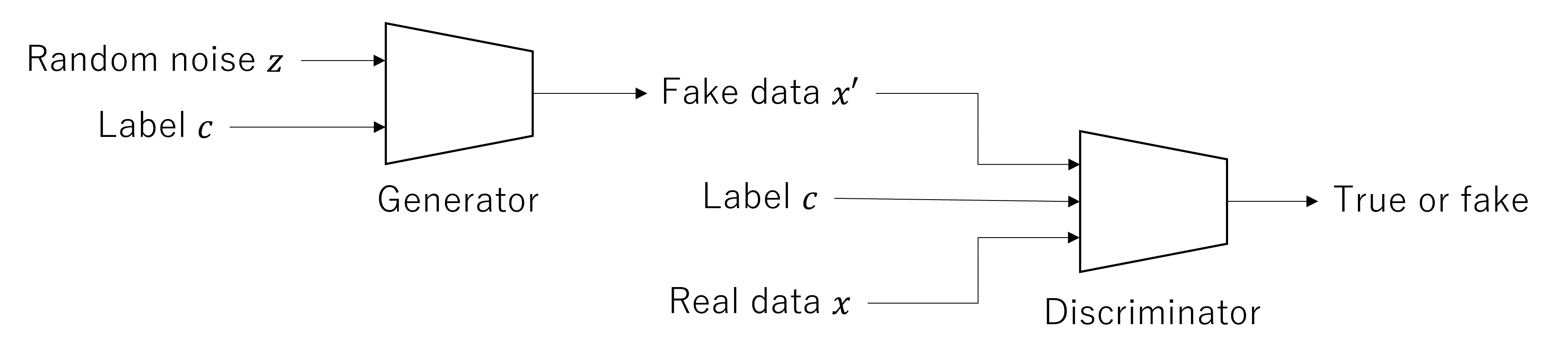}
			\end{center}
		\end{minipage}%
		\caption{Conditional GAN.}
		\label{fig:GAN}
	\end{center}
\end{figure}

GAN \cite{Goodfellow14} consists of a generator network $G$ and discriminator network $D$ as illustrated in \reffig{fig:GAN}. 
The generator generates data that mimics the training data, which are also called real data, and the discriminator distinguishes real data from fake data. 
The generator and  discriminator minimize and maximize the loss function, respectively. 
\begin{align}
	&	\min_G \max_D ~ \mathcal{L}_{GAN} (G, D). 
\end{align}
The loss function is defined as follows: 
\begin{align}
	&	\mathcal{L}_{GAN} (G, D) = E_{\bi{x} \sim p_r(\bi{x})} \left[ \log (D(\bi{x})) \right] + E_{\bi{z}\sim p_z(\bi{z})} \left[ \log \left( 1- D(G(\bi{z})) \right) \right] , \label{eq.LGAN}
\end{align}
where, $\bi{x}$ denotes real data, and $p_r(\bi{x})$ is the probability distribution of the real data. 
$\bi{z}$ denotes a random noise vector sampled from probability distribution $p_z(\bi{z})$. 
$G$ and $D$ denote the generator and discriminator networks, respectively. 
$G(\bi{z})$ is the output from the generator, which is referred to as fake data. 

Although many successful applications of GAN have been reported \cite{Gui21}, 
an instability in training GAN was also reported \cite{Arjovsky17a,goodfellow2017nips}. 
WGAN-gp \cite{Gulrajani} is a solution for improving the stability. 
WGAN-gp uses the earth-mover's (EM) distance instead of the Jensen–Shannon (JS) divergence in GAN. 
EM distance  $W(\cdot , \cdot)$ is a distance between two probability distributions defined by 
\begin{align}
	&	W(p_r, p_g) = \inf_{\gamma \sim \Pi (p_r, p_g)} \mathbb{E}_{\bi{x}, \bi{y} \sim \gamma} [ \| \bi{x} - \bi{y} \| ] ,
\end{align}
where $p_r$ and $p_g$ are probability distributions. 
Then, the loss function of WGAN-gp is 
\begin{align}
	\mathcal{L}_{WGAN{\mathchar`-}gp} (f_w, g_\theta) = W(p_r, p_g) + \lambda \mathbb{E}_{\bi{x}\sim p_r} [ \left( \| \nabla_{\bi{x}} f_w(\bi{x}) \| -1 \right)^2 ], 
\end{align}
where $f_w$ and $g_\theta$ represent the generator and discriminator, respectively.

\subsection{Conditional VAE}
\begin{figure}[tbph]
	\begin{center}
		\begin{minipage}[h]{\textwidth}
			\begin{center}
				\includegraphics[width=\textwidth]{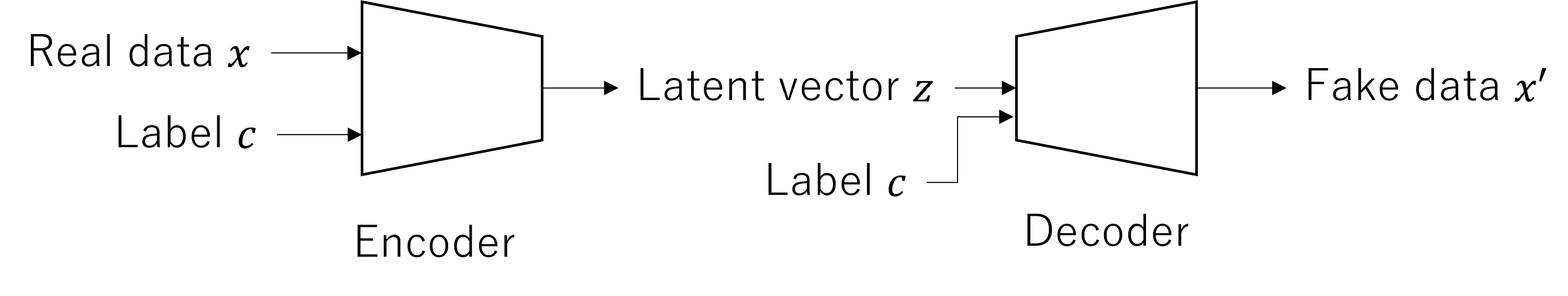}
			\end{center}
		\end{minipage}%
		\caption{Conditional VAE.}
		\label{fig:VAE}
	\end{center}
\end{figure}

The VAE \cite{Kingma13} is an encoder-decoder type deep neural network. 
The VAE model is shown in \reffig{fig:VAE}.  
The encoder embeds features into the latent space, and the decoder reconstructs the data from the latent space; 
\begin{align}
	&	f_{enc} (\bi{x}) = \left( \bi{\mu}, \bi{\sigma^2} \right)^{\top}, \\
	&	\bi{z} \sim N ( \bi{\mu}, \bi{\sigma^2} ), \\
	&	f_{dec} (\bi{z}) = \bi{x}'.
\end{align}
The loss function of the VAE model is 
\begin{align}
	\mathcal{L}_{VAE} &= \mathcal{L}_{\rm llike} + \mathcal{L}_{\rm prior} \\
	&= \mathbb{E}_{\bi{x}} \left( \| \bi{x} - \bi{x}' \|^2 \right)
	+ D_{\rm KL}(q(\bi{z} | \bi{x}) || p(\bi{z}) ),
\end{align}
where $D_{KL} (q(\bi{z} | \bi{x}) || p(\bi{z}) )$ is the Kullback-Leibler (KL) divergence between the distribution of samples in latent space $q(\bi{z} | \bi{x})$ and the prior $p(\bi{z})$. 
The standard normal distribution is used as prior $p(\bi{z})$.

\subsection{Conditional VAEGAN}
\begin{figure}[tbph]
	\begin{center}
		\begin{minipage}[h]{\textwidth}
			\begin{center}
				\includegraphics[width=\textwidth]{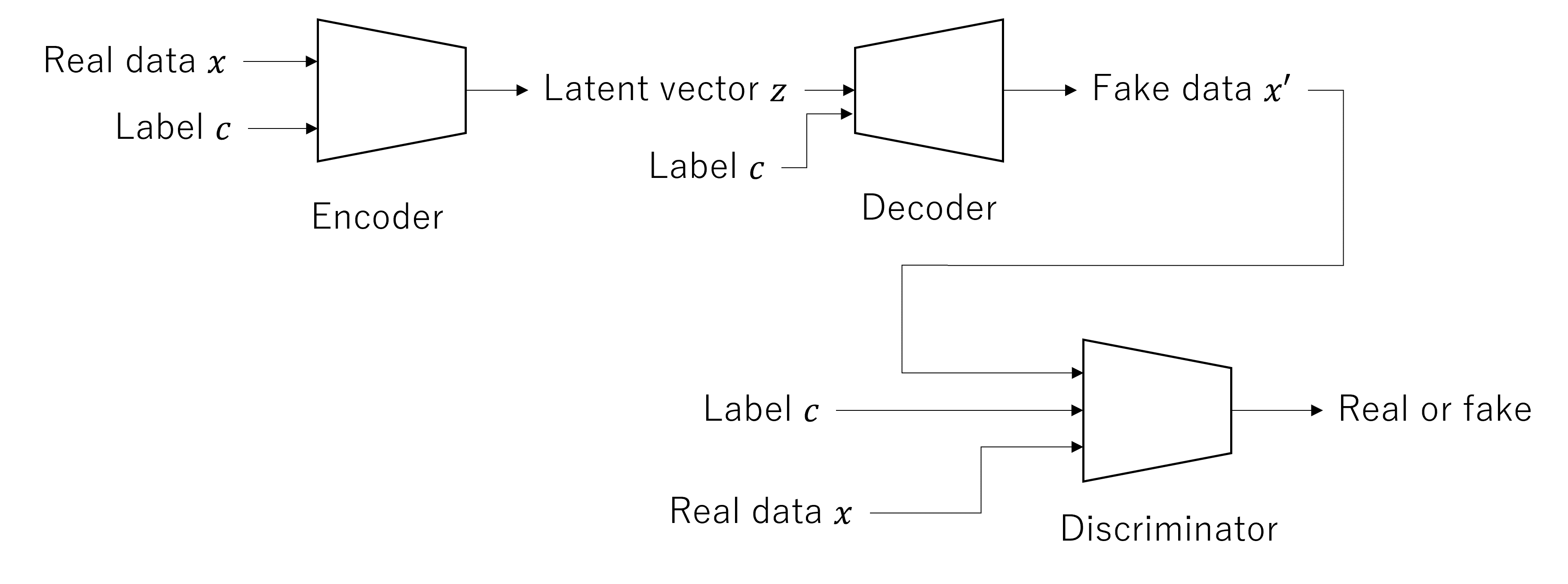}
			\end{center}
		\end{minipage}%
		\caption{Conditional VAEGAN.}
		\label{fig:VAEGAN}
	\end{center}
\end{figure}

A VAEGAN couples the VAE  and GAN models, and it consists of an encoder, decoder, and discriminator, as illustrated in \reffig{fig:VAEGAN}. 
The encoder extracts features from the data and embeds them into the latent space, whereas the decoder generates data from latent vectors.
When generating new data in the VAEGAN, latent vectors are specified and the decoder is processed from the latent vectors. 
The loss function of the VAEGAN model is 
\begin{align}
	\mathcal{L}_{\rm VAEGAN} = \mathcal{L}^{\rm Dis}_{\rm llike} + \mathcal{L}_{\rm GAN} + \mathcal{L}_{\rm prior} .
\end{align}
$\mathcal{L}^{\rm Dis}_{\rm llike}$ denotes the reconstruction loss represented by the discriminator. 
$\mathcal{L}_{\rm GAN}$ and $\mathcal{L}_{\rm prior}$ are the GAN loss and prior loss defined in the GAN and VAE models, respectively.

\section{CVAE -WGAN-gp model for airfoil generation}

\subsection{Conditional VAE-WGAN-gp}
VAEGAN is coupled with WGAN-gp to improve stability and is trained in a conditional manner. 
The loss function of VAE-WGAN-gp is expressed as
\begin{align}
	\mathcal{L}_{\rm VAE{\mathchar`-}WGAN{\mathchar`-}gp} = \mathcal{L}^{\rm Dis}_{\rm llike} + \mathcal{L}_{\rm WGAN{\mathchar`-}gp} + \mathcal{L}_{\rm prior} .
\end{align}
The architecture of the model is the same as that in \reffig{fig:VAEGAN}. 
The training dataset was fed to the encoder with labels, and the data were embedded into the latent space. 
The prior of the latent space was employed as the standard normal distribution. 
The decoder reconstructs data from the latent vector and labels. 
The generated fake data, $\bi{x}'$, were fed into the discriminator with the label and training data. 
The discriminator distinguishes between real data and fake data. 
When generating new data, the latent vector is inputted into the decoder, and the decoder outputs new data. 

The encoder consisted of five layers with 512, 256, 128, and 64 nodes. 
The decoder also consisted of five layers with 64, 128, 256, 512, and 496 nodes. 
The discriminator consisted of three layers with 512, 256, and one node. 
A leaky rectified linear unit (leaky ReLU) \cite{LRELU} was employed as the activation function. 
The dimension of the latent space was grid-searched, and $4 $ was chosen. 
The Adam optimizer was used, the learning rate was set as $0.0001$, and the model was trained for $50,000$ epochs. 

\subsection{Airfoil generation framework}
First, a training dataset was prepared. The dataset consisted of shape data and the corresponding performance indices, which are described in section \ref{sec:dataset}. 
The CVAE-WGAN-gp model was trained using this dataset . 
Once the training was completed, the decoder was used to generate new data. 
The decoder inputs latent vectors and labels and outputs new data. 
When generating the data, latent vectors were sampled using a standard normal distribution. 
The aerodynamic performance of the generated data was evaluated. In the numerical experiments, XFoil \cite{XFOIL} was used to evaluate airfoils.
The aim of the inverse design is to obtain shapes that satisfy the required labels. 
However, the loss function of the CVAE-WGAN-gp model does not consider the reproduction of aerodynamic performance. 
Therefore, the reproduction of the aerodynamic performance is not guaranteed by the model. 
To evaluate the model, the error in aerodynamic performance must be evaluated.

\section{Numerical experiments}
\subsection{Dataset}\label{sec:dataset}
The airfoil dataset was constructed using the NACA 4-digit airfoil dataset \cite{Abbot}. 
The 4-digit airfoil is defined by three parameters: maximum camber, position of maximum camber, and maximum thickness normalized by chord length. 
The number of airfoils generated for the training dataset was $3709$. 
Moreover, the lift coefficient was calculated for each airfoil using XFoil which is a software based on the panel method. 
The airfoil shape is represented by a set of points, where the number of points is $248$ because the XFoil calculation requires more than 120 points. 
$x$ and $y$ coordinates of the points were assembled into one vector as $\bi{x} = \left( x_1, x_2, \dots, x_{248}, y_1, y_2, \dots , y_{248} \right) $.

\begin{figure}[tbph]
	\begin{center}
		\begin{minipage}[h]{\textwidth}
			\begin{center}
				\includegraphics[width=0.5\textwidth]{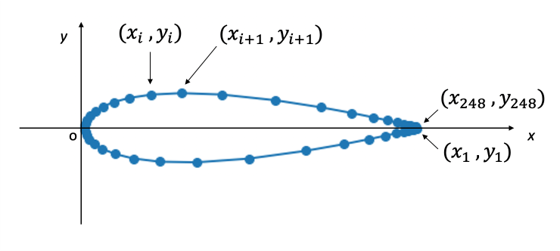}
			\end{center}
		\end{minipage}%
		\caption{Shape discretization.}
		\label{fig:shape}
	\end{center}
\end{figure}

\subsection{Airfoil generation}
\begin{figure}[tbph]
	\begin{center}
		\begin{minipage}[h]{\textwidth}
			\begin{center}
				\scalebox{0.7}{
					\includegraphics{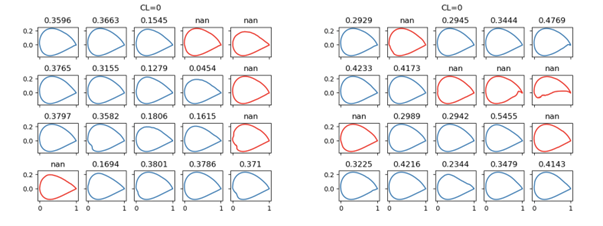}
				}\par
				{\footnotesize (a) $C_{\rm L}=0.0$ (left: CVAE-WGAN-gp, right: cWGAN-gp). }
			\end{center}
		\end{minipage}%
		\par
		\begin{minipage}[h]{\textwidth}
			\begin{center}
				\scalebox{0.7}{
					\includegraphics{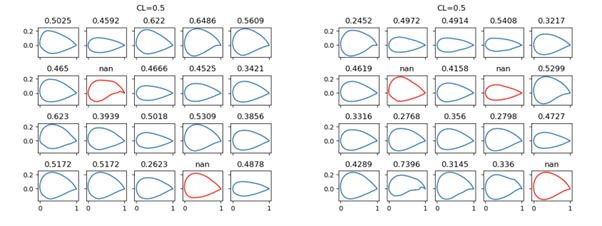}
				}\par
				{\footnotesize (b) $C_{\rm L}=0.5$ (left: CVAE-WGAN-gp, right: cWGAN-gp). }
			\end{center}
		\end{minipage}%
		\par
		\begin{minipage}[h]{\textwidth}
			\begin{center}
				\scalebox{0.7}{
					\includegraphics{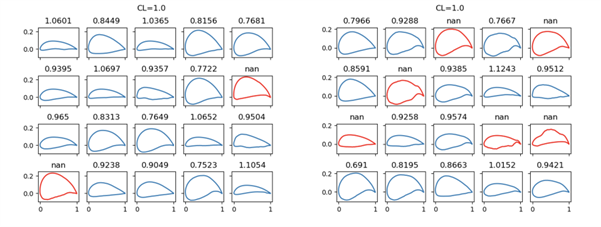}
				}\par
				{\footnotesize (c) $C_{\rm L}=1.0$ (left: CVAE-WGAN-gp, right: cWGAN-gp). }
			\end{center}
		\end{minipage}%
		\par
		\begin{minipage}[h]{\textwidth}
			\begin{center}
				\scalebox{0.7}{
					\includegraphics{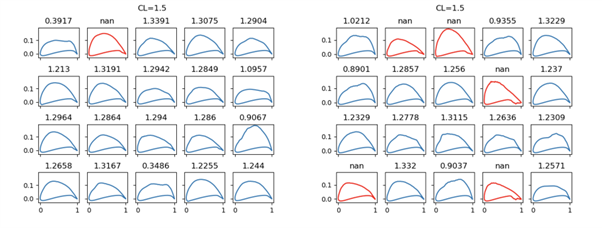}
				}\par
				{\footnotesize (d) $C_{\rm L}=1.5$ (left: CVAE-WGAN-gp, right: cWGAN-gp). }
			\end{center}
		\end{minipage}%
		\caption{Generated shapes (red: not converged, blue: converged).}
		\label{fig:shapes}
	\end{center}
\end{figure}

The proposed CVAE-WGAN-gp and conventional cWGAN-gp models were trained, and the generated shapes are shown in \reffig{fig:shapes}. 
The red figures indicate that the XFoil calculations for the shapes did not converge, whereas the blue figures indicate otherwise. 
Some of the generated shapes are similar to NACA airfoils, and others are wired from an aeronautical point of view. 
To evaluate the accuracy of shape generation quantitatively, the following indices were calculated: 
\begin{itemize}
	\item $\phi_{\rm mean}$: Smoothness index. 
	\item MSE: Mean squared error of $C_{\rm L}$. 
	\item $\mu$: Index of variety of generated shapes.
\end{itemize}
The success, failure, and non-convergence rates provides qualitative indications of accuracy. 

The smoothness index $\phi_{\rm mean}$ is the mean of the smoothness indices $\phi$ defined for each shape. 
$\phi$ is defined as 
\begin{align}
	&	\bi{v}_k = \left( \bi{x}_{k+1} - \bi{x}_{k}, \bi{y}_{k+1} - \bi{y}_{k} \right)^{\top} \\
	&	\phi = \sum_{k=1}^{N} \arccos \left( \frac{ {\bi{v}_k}^{\top} \bi{v}_{k+1} }{ \| {\bi{v}_k} \| \| {\bi{v}_{k+1}}\|  } \right) 
\end{align}
If a shape is a complete circle, $\phi=2 \pi$. If a shape contains zigzag lines, $\phi$ indicates a larger value. 
$\phi_{\rm mean}$ indicates the smoothness of the generated shapes, and a smaller value indicates better smoothness.  
$\phi_{\rm mean}$ of the training dataset is $\phi_{\rm mean} = 2.14 \pi$. The number is slightly larger than $2 \pi$ because of the camber line of the airfoils; the centerline of many airfoils is an upward convex curve. 

The MSE is a quantitative index of the lift coefficient error. 
The MSE is defined as the mean of $\| C_{\rm L}^{\rm r} - C_{\rm L}^{\rm l} \|^2 $ for all shapes whose XFoil calculation converges. 
This MSE is not included in the loss function because the reconstruction error in the loss function is the distance between the output shape and the training shape, whereas this MSE is the distance between the lift coefficients. 
$\mu$ is defined as the mean deviation of shapes from the mean shape, that is, $\| \bi{g}_i - \bi{g}_{\rm mean} \| $, where $\bi{g}_i$ is a vector of a shape and $\bi{g}_{\rm mean}$ is the mean of $\bi{g}_i$. 

The indices of the proposed CVAE-WGAN-gp model and other models are compared in \reftab{tab:success}. 
cWGAN-gp and CVAE-WGAN-gp achieved better scores than the others with respect to $\phi_{\rm mean}$. 
cWGAN-gp and CVAE indicated better scores only for one or two indices, that is, cWGAN-gp was better in $\phi_{\rm mean}$ and $\mu$, and CVAE was better in MSE. However, CVAE-WGAN-gp showed good scores for all the three indices. 

It is reasonable that the MSE of CVAE-WGAN-gp and CVAE are almost similar because both models use the loss between the training data and generated data, that is, $\mathcal{L}_{\rm llike}$ and $\mathcal{L}_{\rm llike}^{\rm Dis}$, which does not contain the loss of cWGAN-gp. Because loss $\mathcal{L}_{\rm llike}$ measures the direct distance between the training and generated data, the model can generate data similar to the training data, which leads to a lower MSE. 

$\phi_{\rm mean}$ was smaller in the cWGAN-gp and CVAE-WGAN-gp models than in the CVAE model. 
A smaller $\phi_{\rm mean}$ was obtained owing to the Wasserstein distance, as shown  in the difference between cGAN and cWGAN-gp. 
The discriminator measures the distance between the training data and generated data using the Wasserstein distance. 

$\mu$ was significantly large in cWGAN-gp, and CVAE-WGAN-gp followed cWGAN-gp. 
In cWGAN-gp, some of the generated shapes were wired, as shown in \reffig{fig:shapes}. The lift coefficients cannot be calculated for these wired shapes, as illustrated in red in \reffig{fig:shapes}, which leads to a larger MSE value. 
CVAE-WGAN-gp indicated a larger $\mu$ than CVAE, whereas the MSEs of both models were almost equal. 

Therefore, CVAE-WGAN-gp exhibited good properties of both CVAE and cWGAN-gp, owing to its architecture and loss functions. 

\begin{table}[tpb]
	\footnotesize
	\begin{center}
		\caption{Success rates and errors of generated shapes.}
		\label{tab:success}       
		\begin{tabular}{lrrr}
			\hline\noalign{\smallskip}
			& $\phi_{\rm mean} \downarrow$ & MSE$\downarrow$ & $\mu \uparrow$  \\
			\noalign{\smallskip}\hline\noalign{\smallskip}
			cGAN $(d=3)$ 
			& 4.91$\pi$ & 0.047 & 0.152\\
			cWGAN-gp $(d=3)$ 
			& ${3.46\pi}$ &0.047 & 0.320 \\
			$\mathcal{N}$-CVAE \cite{Yonekura21a} 
			& 3.95$\pi$ & 0.027 & 0.226\\
			VAE-WGAN-gp $(d=4)$ 
			& 3.50$\pi$ & 0.028 & 0.243\\
			\noalign{\smallskip}\hline& 
		\end{tabular}%
	\end{center}%
\end{table}%

\subsection{Latent distribution}
\begin{figure}[tbph]
	\begin{center}
		\begin{minipage}[h]{0.49\textwidth}
			\begin{center}
				\scalebox{0.5}{
					\includegraphics{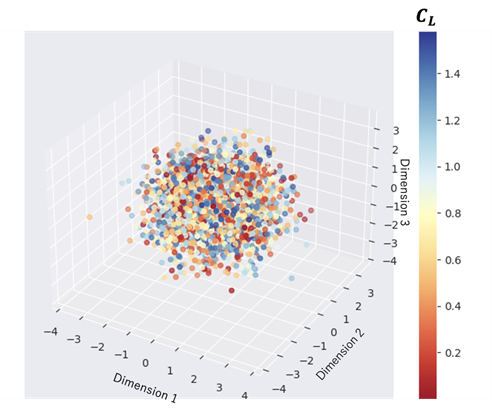}
				}
				\par
				{\footnotesize (a) cWGAN-gp. }
			\end{center}
		\end{minipage}%
		\begin{minipage}[h]{0.49\textwidth}
			\begin{center}
				\scalebox{0.5}{
					\includegraphics{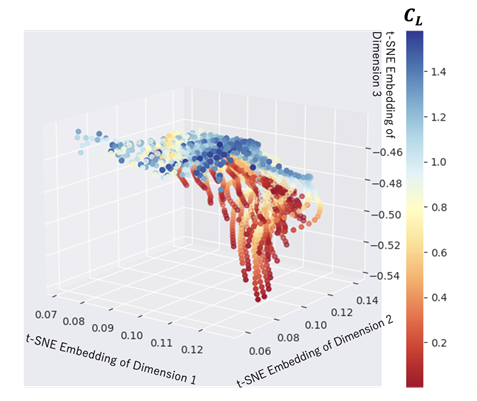}
				}\par
				{\footnotesize (b) CVAE-WGAN-gp. }
			\end{center}
		\end{minipage}%
		\caption{Latent distribution.}
		\label{fig:latent_cWGAN-gp}
	\end{center}
\end{figure}

The latent distributions of cWGAN-gp and CVAE-WGAN-gp are shown in \reffig{fig:latent_cWGAN-gp}. 
Because  the dimension of the latent space of CVAE-WGAN-gp was four , the dimension was reduced using the t-SNE method \cite{tsne} for visualization. 
The latent distribution of cWGAN-gp was not ordered, and it was randomly distributed. 
However, the latent distribution of the CVAE-WGAN-gp model was ordered and lined up neatly; data with low $C_{\rm L}$ and those with high $C_{\rm L}$ are clustered. 
This difference arises from the encoder part of CVAE-WGAN-gp. 

The generator or decoder generates new data from latent data. 
In cWGAN-gp, the generator generates data with both high $C_{\rm L}$ and low $C_{\rm L}$ from a similar latent variable. 
Because the label is also an input to the generator as well as the latent variable, the generator can output different shapes from the same latent variable. However, it is desirable if different data are located in different areas of the latent space. 
In the CVAE-WGAN-gp model, the decoder can generate different data from different latent variables, which results in a lower MSE than cWGAN-gp. 

\section{Conclusion}
This study proposed CVAE-WGAN-gp, a model that generates an airfoil that satisfies required lift coefficients. 
The proposed model showed a better performance in terms of smoothness, reproduction of lift coefficients, and shape varieties, whereas CVAE and cWGAN-gp exhibited some but not all these three desired properties. The proposed model has properties of both the CVAE and cWGAN-gp models. 
The latent space was compared with both the proposed and cWGAN-gp models, and it was shown that the encoder of the CVAE-WGAN-gp model embeds data in the latent space in a neat manner, and the latent data are ordered in the latent space. 
A better MSE is explained by the neatness of the latent space. The fact that different data are embedded in different latent areas helps the decoder to generate accurate data from the latent space. 

\section*{Acknowledgement}
This work was supported by JSPS KAKENHI Grant Numbers JP21K14064 and  JP23K13239.









\end{document}